\documentclass[a4paper,11pt]{article}
\pdfoutput=1 

\usepackage{jcappub} 

\usepackage{graphicx}
\usepackage{bm}
\usepackage{amsmath}
\usepackage{amssymb}
\usepackage{xcolor}
\usepackage{hyperref}
\usepackage[capitalise]{cleveref}
\usepackage[activate={true,nocompatibility},final,kerning=true,factor=1100,stretch=10,shrink=10]{microtype}
\usepackage{academicons}
\usepackage{xcolor}
\usepackage{fontawesome5}

\definecolor{orcidlogocol}{rgb}{0.65, 0.807, 0.223}

\newcommand{\orcid}[1]{$\,$\href{https://orcid.org/#1}{\textcolor{orcidlogocol}{\faOrcid}}}

\usepackage[T1]{fontenc} 

\title{\boldmath Effect of clustering on primordial black hole microlensing constraints}

 \author{Matthew Gorton \orcid{0000-0001-9360-672X}}
  \author{and Anne M. Green \orcid{0000-0002-7135-1671}}
  \affiliation{School of Physics and Astronomy, \\ University of Nottingham
 \\Nottingham NG7 2RD, United Kingdom}

\emailAdd{matthew.gorton@nottingham.ac.uk}
\emailAdd{anne.green@nottingham.ac.uk} 

\abstract{Stellar microlensing observations tightly constrain compact object dark matter in the mass range $(10^{-11} - 10^{3}) M_{\odot}$. Primordial Black Holes (PBHs) form clusters, and it has been argued that these microlensing constraints are consequently weakened or evaded. For the most commonly studied PBH formation mechanism, the collapse of large gaussian curvature perturbations generated by inflation, the clusters are sufficiently extended that the PBHs within them act as individual lenses. We find that if the typical mass of the clusters is sufficiently large, $ \gtrsim 10^{6} M_{\odot}$, then the event duration distribution can deviate significantly from that produced by a smooth dark matter distribution, in particular at the shortest durations. As a consequence of this, the probability distribution of the number of observed events is non-Poissonian, peaking at a lower value, with an extended tail to large numbers of events. However, for PBHs formed from the collapse of large inflationary perturbations, the typical cluster is expected to contain $\sim 10^{3}$ PBHs. In this case the effect of clustering is negligibly small, apart from for the most massive PBHs probed by decade-long stellar microlensing surveys ($M_{\rm PBH} \sim 10^{3} M_{\odot}$).}

\begin{document}
\maketitle
\flushbottom

\section{\label{sec:intro}Introduction}

The discovery of gravitational waves from mergers of tens of Solar mass black holes by LIGO-Virgo~\cite{LIGOScientific:2016aoc} has led to increased interest in Primordial Black Holes (PBHs) as a dark matter (DM) candidate~\cite{Bird:2016dcv,Clesse:2016vqa,Sasaki:2016jop,Carr:2016drx}. PBHs are black holes that may form in the early Universe~\cite{zn,Hawking:1971ei}. The most commonly studied formation mechanism is the collapse of large density perturbations produced by inflation (for reviews see e.g. Refs.~\cite{Carr:2020xqk,Green:2020jor}). 


Stellar microlensing is the temporary amplification which occurs when a compact object passes close to the line of sight to a star~\cite{Paczynski:1985jf}. Various microlensing surveys have placed tight constraints on the abundance of compact objects in the Milky Way (MW) halo.
The OGLE Galactic bulge survey~\cite{Niikura:2019kqi} and observations of M31 using Subaru HSC~\cite{Niikura:2017zjd,Croon:2020ouk} constrain planetary and sub-planetary masses, while the EROS~\cite{EROS-2:2006ryy}, MACHO~\cite{Macho:2000nvd} and OGLE~\cite{2016MNRAS.458.3012W} surveys of the Large and Small Magellanic Clouds (LMC and SMC) constrain stellar and planetary masses. Following a proposal by Ref.~\cite{2018A&A...618L...4M}, Ref.~\cite{2022arXiv220213819B} has combined data from EROS-2 and MACHO to obtain sensitivity to long duration events, and hence constrain more massive compact objects. Taken at face value, the stellar microlensing constraints exclude PBHs with mass $ 10^{-11} M_{\odot} \lesssim M_{\rm PBH} \lesssim 10^{3} M_{\odot}$ making up all of the DM. However the calculation of these constraints involves various assumptions, for instance that the DM is smoothly distributed.

PBHs that form from the collapse of large gaussian perturbations generated by inflation do not form in gravitationally bound clusters~\cite{Ali-Haimoud:2018dau}. However, since PBHs are discrete objects, there are Poisson fluctuations in their initial distribution. As a consequence of these isocurvature fluctuations in the PBH density, PBH clusters form not long after radiation-matter equality~\cite{Afshordi:2003zb}. The abundance and properties of these clusters have been studied numerically~\cite{Inman:2019wvr} and analytically~\cite{Jedamzik:2020ypm}, using the spherical top-hat collapse model.
Refs.~\cite{Garcia-Bellido:2017xvr,Calcino:2018mwh,Carr:2019kxo} have argued that PBH clustering modifies the stellar microlensing constraints so that they are shifted to lower masses, and consequently multi-Solar mass PBHs can make up all of the DM.

In this manuscript we examine the effect of clustering on the stellar microlensing constraints for the clusters which form when PBHs are produced by the collapse of large gaussian inflationary density perturbations. In Sec.~\ref{sec:cluster} we overview the properties of the PBH clusters. Next, in Sec.~\ref{sec:eventrate}, we outline the calculation of the microlensing differential event rate, firstly for the standard case of a smooth MW halo (Sec.~\ref{sec:smooth}) and then for clustered DM (Sec.~\ref{sec:clustered}). We present our results in Sec.~\ref{sec:results} before concluding with discussion in Sec.~\ref{sec:discuss}.

\section{Cluster properties}
\label{sec:cluster}

Jedamzik used the spherical top-hat collapse model to calculate the properties of the PBH clusters that form when PBHs generated from the collapse of large inflationary density perturbations, with a single mass, make up all of the dark matter~\cite{Jedamzik:2020ypm}. The initial fluctuation in the density in a region containing $N$ PBHs is $\delta(N)= 1/\sqrt{N}$, and these isocurvature fluctuations grow with time proportional to \cite{Afshordi:2003zb}  \begin{equation}
D(a) \approx \left( 1 + \frac{3 a}{2 a_{\rm eq}} \right) \,,
\end{equation}
where $a_{\rm eq}$ is the scale factor at radiation-matter equality.
A particular scale goes non-linear when the scale factor is equal to $a_{\rm coll}$, determined by $D(a_{\rm coll}) \delta(N) \approx \delta_{\rm c} \approx 1.68$~\footnote{The threshold for collapse is in fact slightly larger than the standard value of 1.68 for scales which collapse not long after radiation-matter equality, i.e.~if $a_{\rm coll} \sim a_{\rm eq}$ (see Appendix A and Fig. 14 of Ref.~\cite{Inman:2019wvr}). However this does not have a significant effect on the estimates of the cluster properties.}. The resulting gravitationally bound cluster has density approximately $178$ times the background dark matter density at this time: $\rho_{\rm cl} \approx 178 \rho_{\rm dm}(a_{\rm coll})$. The number density, $n_{\rm cl}$, of PBHs within a cluster containing $N_{\rm cl}$ PBHs is then 
\begin{equation}
n_{\rm cl} = \frac{\rho_{\rm cl}}{M_{\rm PBH}} \approx 1. 7 \times 10^{5} \,  N_{\rm cl}^{-3/2} \left( \frac{M_{\odot}}{M_{\rm PBH}} \right)  \, {\rm pc}^{-3} \,,
\end{equation}
and the cluster radius, $R_{\rm cl}$, can be estimated, from $(4 \pi/3) n_{\rm cl} R_{\rm cl}^3 = N_{\rm cl}$, to be
\begin{equation}
\label{rcl}
R_{\rm cl} \approx 1.1 \times 10^{-2} \, N_{\rm cl}^{5/6} \left( \frac{M_{\rm PBH}}{M_{\odot}} \right)^{1/3}  \, {\rm pc} \,.
\end{equation}

For initially Poisson distributed discrete objects, the number of clusters containing $N_{\rm cl}$ objects, $\tilde{N}$, is given (for $N_{\rm cl} \gg 1$) by~\cite{1983MNRAS.205..207E,Inman:2019wvr}
\begin{equation}
\tilde{N} \propto \frac{\delta_{\star}}{N_{\rm cl}^{3/2}} \exp{\left(- \frac{N_{\rm cl}}{N_{\star}} \right)} \,,
\end{equation}
where $\delta_{\star}(a) = \delta_{\rm c}/D(a)$
and
\begin{equation}
N_{\star} = \left[ \log{(1+ \delta_{\star})} - \frac{\delta_{\star}}{1+ \delta_{\star}} \right]^{-1} \,.
\end{equation}
$\tilde{N}$ is always a monotonically decreasing function of $N_{\rm cl}$, and $N_{\star}$ grows with time. Therefore clusters containing a small number of PBHs are always the most abundant, however the number of clusters with large $N_{\rm cl}$ increases with time. This behaviour has been confirmed numerically~\cite{Inman:2019wvr}. 

Clusters containing small numbers of objects evaporate~\cite{b+t}, and PBH clusters with $N_{\rm cl} \lesssim 10^{3}$ will have evaporated by the present day~\cite{Afshordi:2003zb,Jedamzik:2020ypm}. 
Therefore, for PBHs that form from large inflationary density perturbations, the most common cluster size today is expected to be $N_{\rm cl} \sim10^{3}$, independent of the PBH mass.

\section{Microlensing event rate}
\label{sec:eventrate}
In this section we outline the calculation of the microlensing differential event rate, first for the standard case of a smooth halo (Sec.~\ref{sec:smooth}) and then for clustered DM (Sec.~\ref{sec:clustered}).

\subsection{Smooth halo}
\label{sec:smooth}

The microlensing differential event rate, ${\rm d} \Gamma/ {\rm d} \hat{t}$, towards the LMC for a smooth halo composed entirely of compact objects with mass $M_{\rm PBH}$ and a Maxwellian velocity distribution is given by~\cite{Griest:1990vu,MACHO:1996qam}:  
\begin{equation}
\label{df}
\frac{{\rm d} \Gamma}{{\rm d} \hat{t}} =  \frac{32 L}
                 { M_{\rm PBH} {\hat{t}}^4
              {v_{{\rm c}}}^2}
              \int^{1}_{0} \rho(x) R^{4}_{{\rm E}}(x)
              e^{-Q(x)}  {\rm d} x \,, 
\end{equation}
where $\hat{t}$ is the time taken to cross the Einstein {\it diameter},  $R_{\rm E}(x)$ is the Einstein radius
\begin{eqnarray}
R_{{\rm E}}(x)&=& 2 \left[ \frac{ G M_{\rm PBH} x (1-x)L}{c^2 } \right]^{1/2} \,, \\
 & \approx & 10^{-4} \, {\rm pc} \left[ \left(\frac{M_{\rm PBH}}{M_{\odot}} \right) \left( \frac{L}{50 \, {\rm kpc}} \right) x (1-x) \right]^{1/2} \,, 
\label{re}
\end{eqnarray}
$G$ is the Gravitational constant, $L \approx 50 \, {\rm kpc}$ is the distance to the LMC, $x$ is the distance of the lens from the observer in units of $L$ and $Q(x)= 4 R^{2}_{{\rm E}}(x) / (\hat{t}^{2} v_{{\rm c}}^2)$, where $v_{\rm c}  =220 \, {\rm km \, s}^{-1}$ is the circular speed.  

The standard halo model usually assumed in microlensing studies (`Model S')~\footnote{The best fit values of some of the parameters appearing in this model have changed in recent years, for instance the Solar radius has been measured as $R_{\rm 0} = (8.18 \pm 0.01 \pm 0.02) \, {\rm kpc}$ by the GRAVITY collaboration~\cite{2019A&A...625L..10G}. However, since these changes have a relatively small effect on the microlensing differential event rate compared with changes in the density profile~\cite{Green:2017qoa,Calcino:2018mwh}, we retain the `traditional' parameter values for consistency with past work in this field.}  is a cored isothermal sphere with density profile
\begin{equation}
\rho(R) = \rho_{0} \frac{R_{{\rm c}}^2 + R_{0}^2}{R_{{\rm c}}^2 + R^2} \,,
\label{rhor}
\end{equation}
and local dark matter density $\rho_{0}= 0.0079 M_{\odot} {\rm pc}^{-3}$, core radius $R_{{\rm c}} = 5$ kpc  and
Solar radius $R_{0} = 8.5$ kpc.  The differential event rate, Eq.(\ref{df}), is then given by~\cite{MACHO:1996qam} 
\begin{equation}
\frac{{\rm d} \Gamma}{{\rm d} \hat{t}} = \frac{512 \rho_{0} 
          (R_{{\rm c}}^2 + R_{0}^2) L G^2  M_{\rm PBH} }
             {  {\hat{t}}^4 {v_{{\rm c}}}^2 c^4}
              \int^{1}_{0} \frac{x^2 (1-x)^2}{A + B x + x^2}
            e^{-Q(x) }{\rm d} x \,, 
\end{equation}
where $A=(R^2_{{\rm c}}+ R^2_{0})/L^2$, $B=-2(R_{0}/L) \cos{b}
\cos{l}$ and $b = -32.8^{\circ}$ and $l = 281^{\circ}$ are the galactic
latitude and longitude, respectively, of the LMC.

The expected number of events, $N_{\rm exp}$, is given by
\begin{equation}
N_{{\rm exp}} = E \int_{0}^{\infty} \frac{{\rm d} \Gamma}{{\rm d} \hat{t}}
           \,  \epsilon(\hat{t}) \, {\rm d} \hat{t} \,,
\label{nexp}           
\end{equation}
where $E$ is the exposure in star years and $\epsilon(\hat{t})$ is the detection efficiency i.e.~the probability that a microlensing event that occurs with duration $\hat{t}$ is detected.

\subsection{Clustered halo}
\label{sec:clustered}

The typical separation of PBHs in a cluster is much larger than the Einstein Radius (for $M_{\rm PBH} = 1 M_{\odot}$, $n_{\rm cl}^{-1/3} \sim 10^{-2} N_{\rm cl}^{1/2} \, {\rm pc}$ while $R_{\rm E} \sim 10^{-4} \, {\rm pc}$).
Therefore the individual PBHs act as lenses, and not (as argued in Ref.~\cite{Garcia-Bellido:2017xvr,Calcino:2018mwh}) the cluster as a whole.  Appendix A2 of Ref.~\cite{Carr:2019kxo} argues that even when clusters are sufficiently diffuse that the PBHs act as individual lenses, lensing by the cluster as a whole renders the magnification from lensing by a single PBH unobservable. Their argument, however, relies on a significant underestimate of the Einstein radius (see Appendix~\ref{sec:app} for further details). A fraction of the PBHs may be in binaries~\cite{Nakamura:1997sm,Ali-Haimoud:2017rtz,Jedamzik:2020ypm}. Ref.~\cite{Petac:2022rio} has however shown that the time separation of the lensing events caused by PBHs in a binary would be of order $100 \, {\rm yr}$, and hence the PBHs act as separate, individual lenses.

Our method for calculating the microlensing event rate from clusters is similar to Refs.~\cite{Maoz:1994dx,Metcalf:2006ms}. We assume that the surface area of the LMC is circular, so that microlensing events can be caused by compact objects within a cone with apex at the Earth and base at the LMC. We take the cone half angle to be $\theta=5.2^{\circ}$, to match the $84 \, {\rm deg}^2$ of the LMC monitored by EROS-2. Using Eq.~(\ref{rhor}) for the density profile of the MW, the total mass of DM in this cone is $M_{\rm cone} \approx 9 \times 10^{8} M_{\odot}$. We assume that a fraction $f$ of the DM is in the form of PBHs~\footnote{To make subsequent notation clearer and more concise, we do not use the usual subscript `PBH' for the fraction of the MW halo in PBHs.}, and all PBHs are in clusters containing $N_{\rm cl}$ PBHs with mass $M_{\rm PBH}$.  We saw in Sec.~\ref{sec:cluster} that clusters with $N_{\rm cl} \lesssim 10^{3}$ will have evaporated by the present day, and therefore some (probably quite large) fraction of PBHs will not be clustered today. Therefore assuming that all PBHs are in clusters with $N_{\rm cl} \gtrsim 10^{3} $ provides an upper limit on the actual effect of clustering on the EROS-2 microlensing constraints. As noted by Peta\v c et al.~\cite{Petac:2022rio}, for large $N_{\rm cl}$ and/or $M_{\rm PBH}$ the cluster radii, $R_{\rm cl}$, given by Eq.~(\ref{rcl}) from the spherical top-hat collapse model are unphysically large. In particular they are larger than the typical separation between clusters. Therefore we follow Ref.~\cite{Petac:2022rio} and set $R_{\rm cl} = 10 \, {\rm pc}$.

In order to take into account clusters that lie only partly within the microlensing cone, we simulate clusters within a larger region which is centered on the microlensing cone and has radius at each line of sight distance, $x$, equal to the radius of the microlensing cone plus the cluster radius: $r_{\rm tcone}(x) = x L \tan{\theta}  + R_{\rm cl} $, i.e.~a truncated cone with the narrow end at the Earth.  For each combination of $M_{\rm PBH}$ and $N_{\rm cl}$ we first calculate the average number of clusters within the truncated cone, $N_{\rm tcone}$,
\begin{equation}
N_{\rm tcone} = \frac{f M_{\rm tcone}}{M_{\rm PBH} N_{\rm cl}} 
\,,
\end{equation} 
where $M_{\rm tcone}$ is the mass within the truncated cone. For each realisation we first draw the actual number of clusters from a Poisson distribution. The line-of-sight position, $x_{\rm cl}$, and transverse velocity, $v_{\perp, {\rm cl}}$, of each cluster are generated assuming the cored isothermal sphere density profile, Eq.~(\ref{rhor}), and a Maxwellian velocity distribution with  $v_{\rm c} =220 \, {\rm km \, s}^{-1}$. We also generate a value for the distance of the centre of the cluster from the axis of the microlensing cone such that, at each $x$, the clusters are uniformly distributed within the circular cross-section of the truncated cone.
This distance is then used to calculate $\hat{f}$, the fraction of the cluster within the microlensing cone.

The velocity dispersion of PBHs within a cluster is of order~\cite{Jedamzik:2020ypm}
\begin{equation}
\sigma_{\rm cl} \approx 0.6 \left( \frac{M_{\rm PBH}}{M_{\odot}} \right)^{1/3} N_{\rm cl}^{1/12} \, {\rm km} \, {\rm s}^{-1} \,.
\end{equation}
This is negligible compared with the cluster transverse velocity, and therefore all PBHs within a given cluster will cause microlensing events with the same duration 
\begin{eqnarray}
\label{that1}
\hat{t}_{\rm cl} & = & \frac{2 R_{\rm E}(x_{\rm cl})}{v_{\perp, {\rm cl}}} \,, \\
   & \approx &  300 \left[ \left(\frac{M_{\rm PBH}}{M_{\odot}} \right) \left( \frac{L}{50 \, {\rm kpc}} \right) x_{\rm cl} (1-x_{\rm cl}) \right]^{1/2} \nonumber \\
       && \times  \left( \frac{220 \, {\rm km \, s}^{-1}}{v_{\perp, {\rm cl}}} \right) \, {\rm days} \,.
\label{that2}    
\end{eqnarray}

Next we need to calculate the rate at which lensing events occur for each cluster.
The optical depth is the probability that a star lies within the Einstein radius of a lens (e.g.~Ref.~\cite{Mao:2008kp}). For a cluster which lies entirely within the microlensing cone the optical depth, $\tau_{\rm cl}$, is the product of the lensing cross section ($ \pi R_{\rm E}^2$), the surface number density of lenses and the fraction of the solid angle to the LMC, $\Omega_{\rm LMC}$, covered by the cluster~\cite{Metcalf:2006ms}:
\begin{eqnarray}
\tau_{\rm cl} &=& (\pi R_{\rm E}^2) \frac{M_{\rm cl}/(\pi R_{\rm cl}^2) }{M_{\rm PBH}} \frac{\Omega_{\rm cl}}{\Omega_{\rm LMC}}  \,, \\
  &=& \frac{ M_{\rm cl} \pi R_{\rm E}^2}{M_{\rm PBH} \Omega_{\rm LMC} x_{\rm cl}^2 L^2} \,,
\end{eqnarray}
where $\Omega_{\rm cl} = \pi R_{\rm cl}^2/(x_{\rm cl}^2 L^2)$ is the solid angle subtended by the cluster. In a time ${\rm d} t$ the lensing area swept out by a lens is ${\rm d} A = 2 R_{\rm E} v_{\perp, {\rm cl}} \, {\rm d} t$ and hence the probability of a new microlensing event occurring is
\begin{eqnarray}
{\rm d} \tau_{\rm cl} &=& \frac{M_{\rm cl}/(\pi R_{\rm cl}^2) }{M_{\rm PBH}} \frac{ \Omega_{\rm cl}}{\Omega_{\rm LMC}} {\rm d} A \,, \nonumber \\
   &=& \frac{\tau_{\rm cl}}{\pi R_{\rm E}^2} 2 R_{\rm E} v_{\perp, {\rm cl}} {\rm d} t \,.
\end{eqnarray}
The rate at which microlensing occurs, $\Gamma_{\rm cl} = {\rm d} \tau_{\rm cl} / {\rm d} t$, is therefore
\begin{equation}
\Gamma_{\rm cl} = \frac{2 \hat{f} v_{\perp, {\rm cl}} M_{\rm cl} R_{\rm E} }{M_{\rm PBH} \Omega_{\rm LMC} x_{\rm cl}^2 L^2} \,.
\label{gammac}  
\end{equation}
where $\hat{f}$ is the fraction of the cluster which lies within the microlensing cone.

\begin{figure*}[ht!]
\begin{center}
\includegraphics[width=0.48\textwidth]{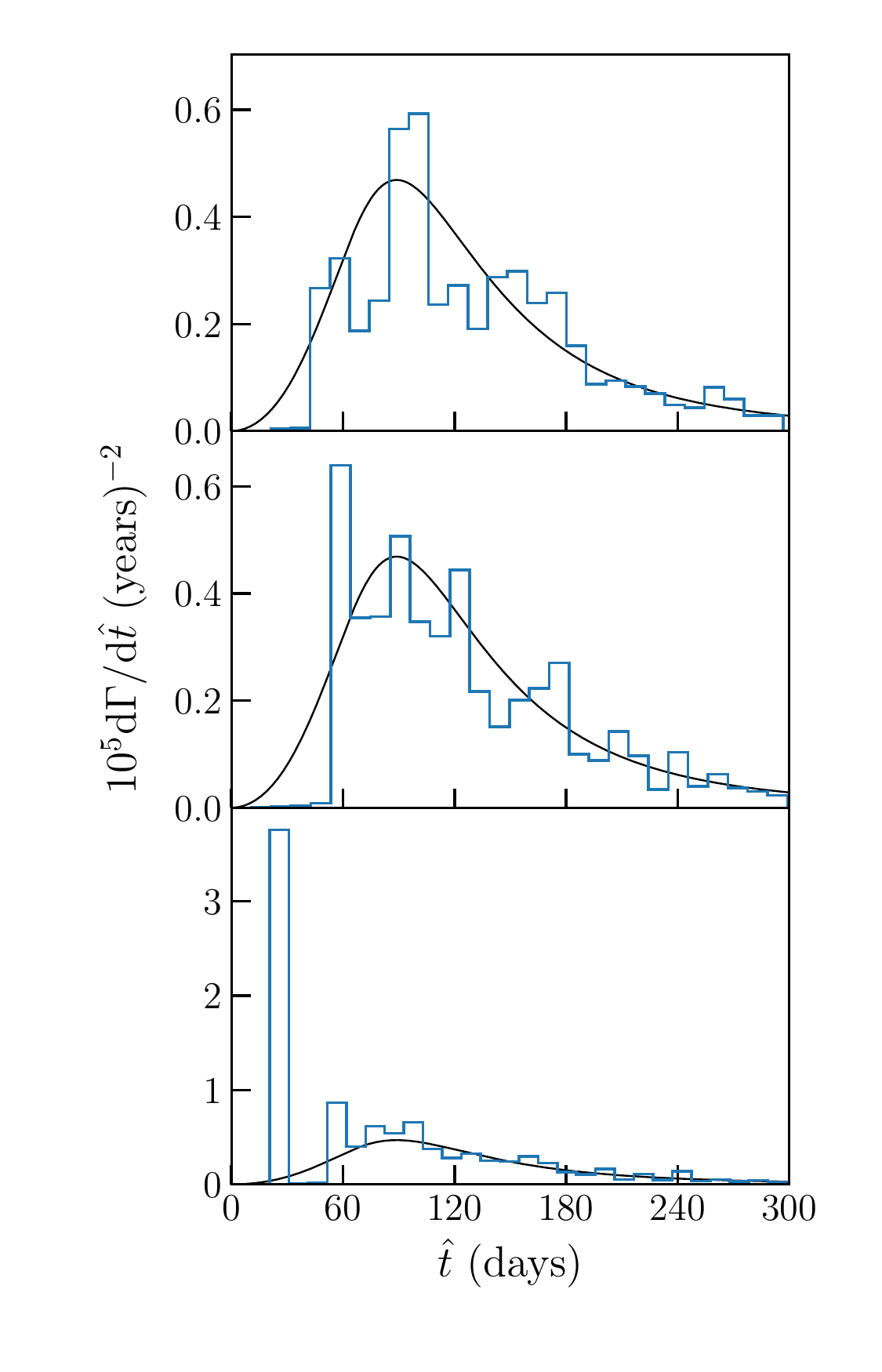}
\includegraphics[width=0.48\textwidth]{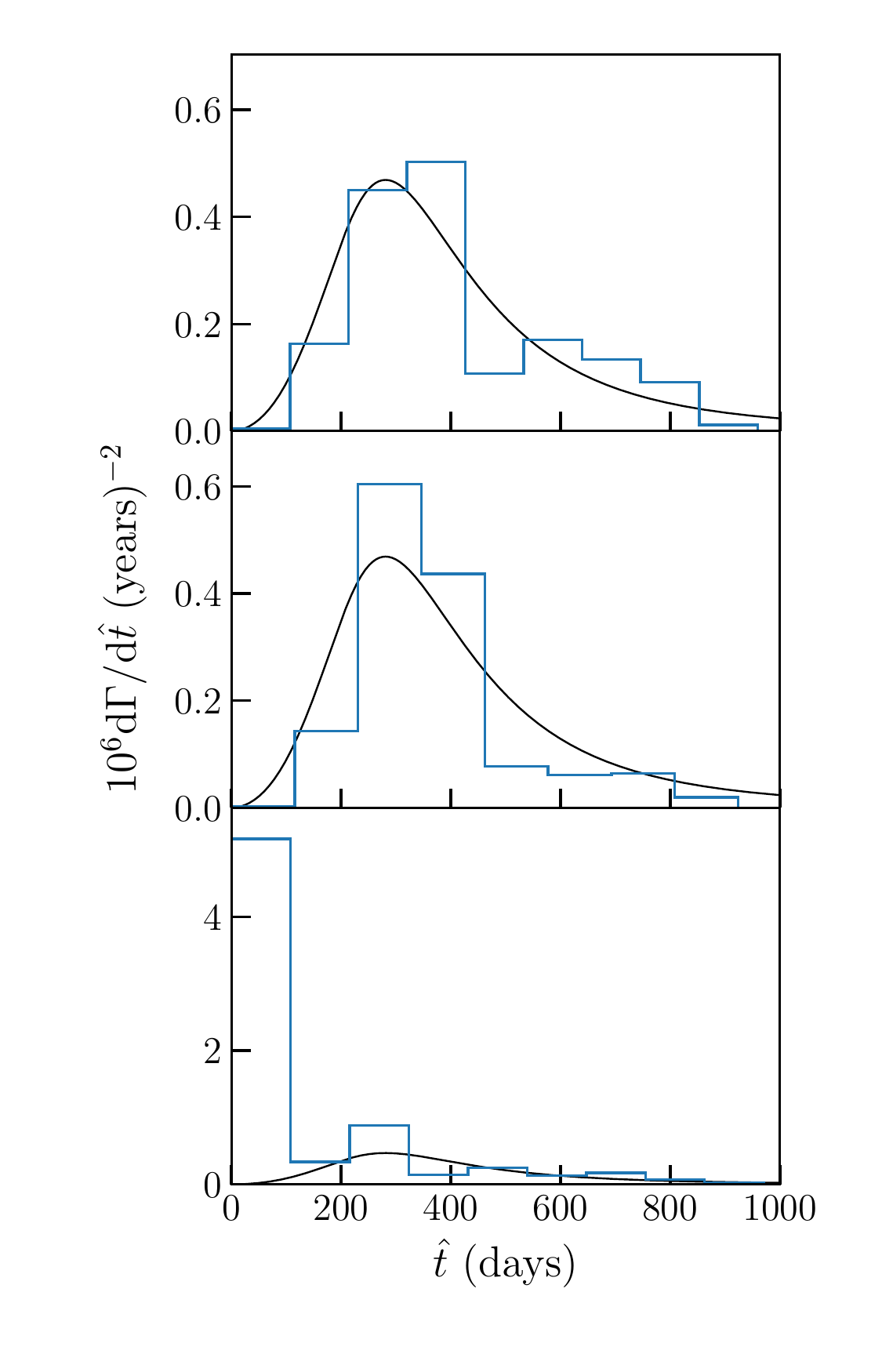}
\end{center}
\caption{Example realisations of the differential event rate, ${\rm d} \Gamma/ {\rm d} \hat{t}$, for clustered DM (blue lines) compared with the standard smooth DM halo (black). In all six cases all of the DM is in clusters containing $N_{\rm cl} = 10^{6}$ PBHs and the PBHs have mass $M_{\rm PBH} = 1$ and $10 M_{\odot}$ in the left and right hand columns respectively. The top two rows show `typical' realisations, where the absence of any clusters close to the observer leads to a deficit of short-duration events. The bottom row shows examples of rare realisations where there is a cluster close to the observer which produces short-duration events at a high rate (note the different range of the y-axis in this case).
\label{fig:eventrate}}
\end{figure*}

For each realisation, we calculate the total differential event rate, ${\rm d} \Gamma/ {\rm d} \hat{t}$, from all clusters by summing the binned values of the event durations, $\hat{t}_{\rm cl}$, for each cluster, weighted by their rates, $\Gamma_{\rm cl}$. 
The mean number of events produced by each cluster is given by
\begin{equation}
\bar{N}_{\rm cl} =  E   \epsilon(\hat{t}_{\rm cl}) \Gamma_{\rm cl} \,.
\end{equation}
For each cluster we draw the observed number of events, $N_{\rm cl, obs}$, from a Poisson distribution with mean $\bar{N}_{\rm cl}$. The total number of observed events, $N_{\rm obs}$, is the sum of $N_{\rm cl, obs}$ over all clusters.


\section{Results}
\label{sec:results}

We use the method described in Sec.~\ref{sec:clustered} to calculate the differential event rate for $10^{4}$ realisations of each combination of the number of PBHs in a cluster, $N_{\rm cl}$, and the PBH mass, $M_{\rm PBH}$.  For cases where the number of clusters in the cone to the LMC is large, $ \gtrsim 10^{4}$, then the differential event rate for a single realisation only has the expected small stochastic deviations from the differential event rate produced by smoothly distributed DM. However when the number of clusters in the cone is smaller than this, there are systematic deviations in the differential event rate for short-duration events. For most realisations there is a deficit of short-duration events, however for a small fraction of realisations there is a large excess of short events.  Fig.~\ref{fig:eventrate} shows the differential event rate, ${\rm d} \Gamma/ {\rm d} \hat{t}$, for three different realisations for $N_{\rm cl} = 10^{6}$ and both $M_{\rm PBH}=1$ and $10 \, M_{\odot}$, compared with the case of smoothly distributed DM. For both values of $M_{\rm PBH}$ we show two `typical' realisations, which have a deficit of short events, and one `rare' realisation with an excess.

\begin{figure*}[t]
\begin{center}
\includegraphics[width=0.48\textwidth]{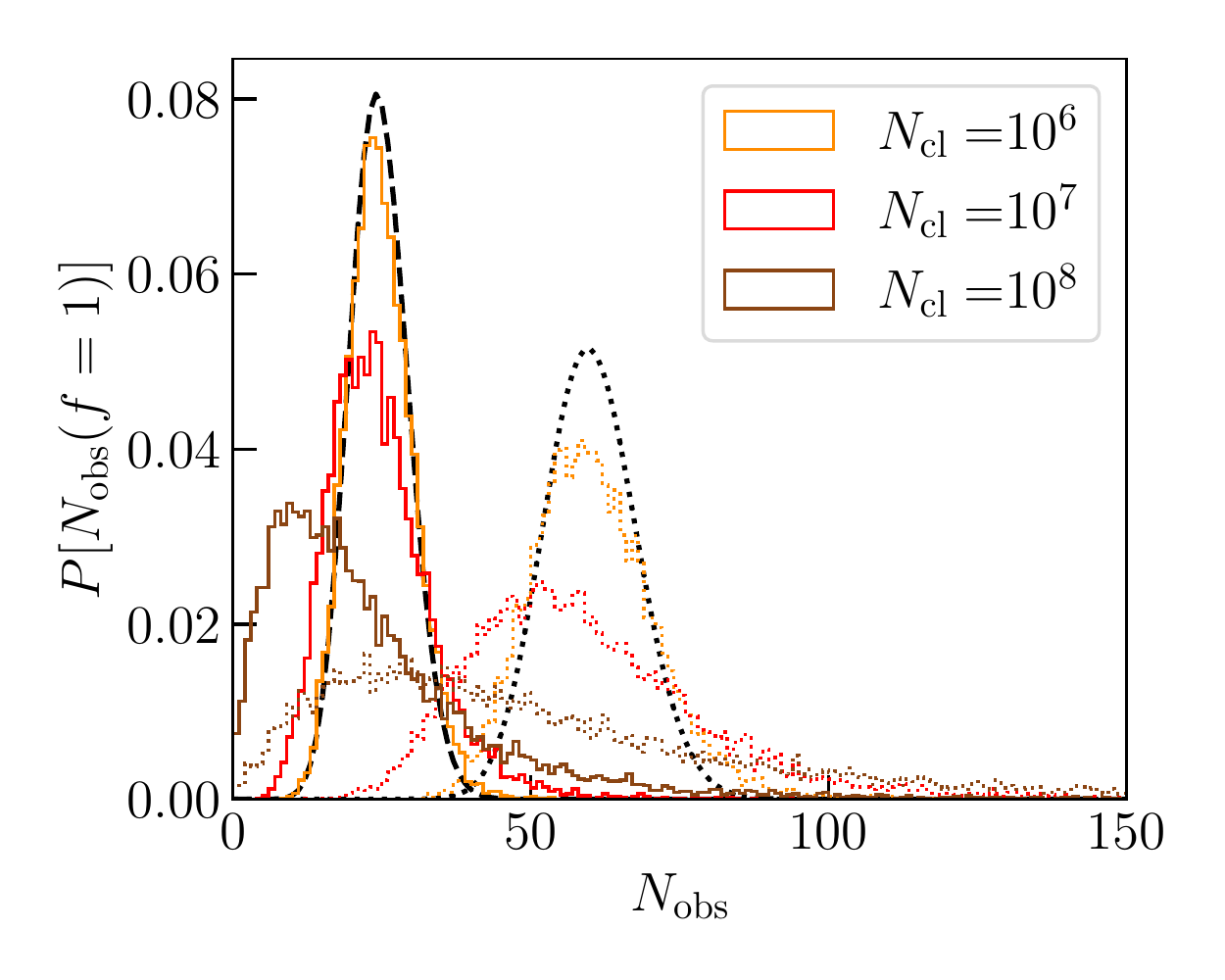}
\includegraphics[width=0.48\textwidth]{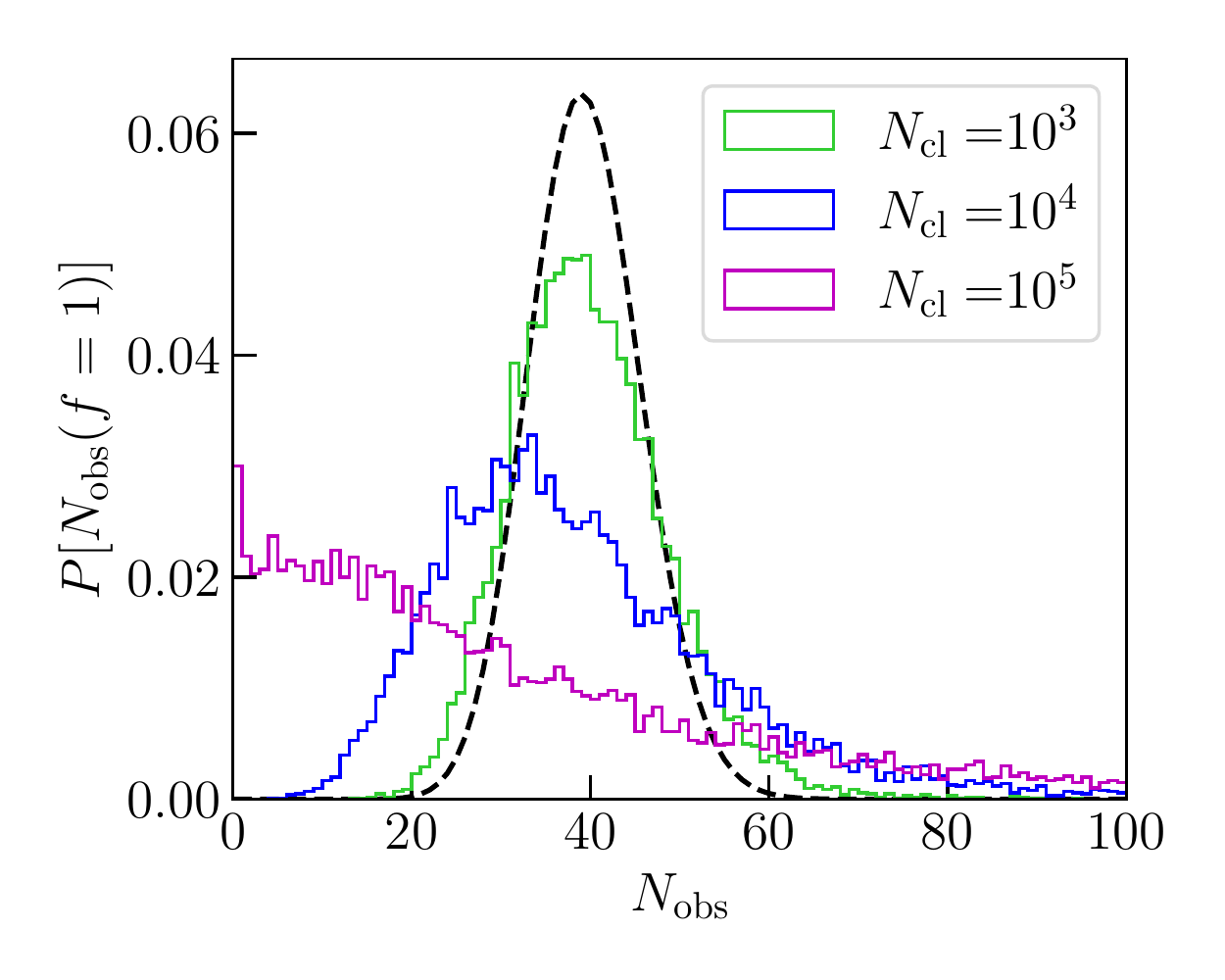}
\end{center}
\caption{The probability distribution, $P[N_{\rm obs}(f=1)]$, of the observed number of microlensing events, $N_{\rm obs}(f=1)$, if all of the MW halo is in compact objects with $M_{\rm PBH}= 1 M_{\odot}$ (left panel) and $M_{\rm PBH}= 10^{3} M_{\odot}$ (right panel). \\
{\it Left panel}:  The orange, red and brown lines assume all of the DM is in clusters with $M_{\rm cl}=10^{6}, 10^{7}$ and $10^{8} M_{\odot}$ respectively. The solid lines use the EROS-2 detection efficiency (see Sec.~\ref{sec:eventrate} for further details) while the dotted lines assume perfect efficiency, i.e.~$\epsilon(\hat{t}) =1$ for all $\hat{t}$.  The black dashed and dotted lines show the Poisson distribution, which arises in the standard case of a smooth DM halo, with $N_{\rm exp}(f=1) = 25$ and $60$ for the EROS-2 and perfect detection efficiencies respectively. \\
{\it Right panel}: The green, blue and purple lines assume all of the DM is in clusters with $M_{\rm cl}=10^{3}, 10^{4}$ and $10^{5} M_{\odot}$ respectively for a `toy' long-duration event survey (see Sec.~\ref{sec:eventrate}). The black dashed line shows the Poisson distribution, which arises in the standard case of a smooth DM halo, which has $N_{\rm exp}(f=1) = 40$.
\label{fig:pnobs}}
\end{figure*}

This behaviour can be understood by considering the dependence of the Einstein radius, $R_{\rm E}$, and the cross-sectional area of the cone to the LMC on $x$, the fractional distance along the line of sight. The Einstein radius is proportional to $[x (1-x)]^{1/2}$ while (for the standard halo model) the lens transverse velocity distribution is independent of $x$. Therefore short-duration events are typically produced by lenses (in the case of clustered DM, clusters) at small or large $x$. The cross-sectional area of the cone to the LMC is proportional to $x^2$, therefore the probability of there being a cluster within (or partly within) the cone at small $x$ is small. Most realisations don't have clusters at very small $x$, and hence have a deficit of short-duration events. For the small fraction of realisations which do have a cluster at very small $x$, that cluster subtends a large fraction of the solid angle to the LMC and hence produces a high rate of short-duration events. More quantitatively, see Eq.~(\ref{gammac}), the total lensing rate by the cluster, $\Gamma_{\rm cl}$, is proportional to $R_{\rm E}/x_{\rm cl}^2$. The `rare' realisations in Fig.~\ref{fig:eventrate} both have a cluster with a small $x_{\rm cl}$ value. 

These variations in the rate of short events emerge when the number of clusters in the cone to the LMC is smaller than $\sim 10^{3}$ (which corresponds to a number of PBHs per cluster $N_{\rm cl} \gtrsim 10^{6} (M_{\odot}/M_{\rm PBH})$) and become larger if the number of clusters is decreased. Since the Einstein radius increases with increasing $M_{\rm PBH}$, so does the value of $\hat{t}$ at which the variations in the differential event rate appear. We note that for the standard PBH formation mechanism, the collapse of large inflationary density perturbations, most clusters are expected to have $N_{\rm cl} \sim 10^{3}$, and not all PBHs are in clusters. Therefore, for this formation mechanism, we expect this effect to be negligible apart from for the most massive PBHs probed by stellar microlensing, $M_{\rm PBH} \sim 10^{3} M_{\odot}$.

Next we study the effect of these variations in the differential event rate on the number of events predicted in LMC microlensing surveys. We consider two different microlensing survey configurations: 
\begin{itemize}
\item An EROS-2-like survey, with exposure $E=3.77 \times 10^{7}$ star years and detection efficiency, $\epsilon(\hat{t})$, given in Fig.~11 of Ref.~\cite{EROS-2:2006ryy}, which observes no microlensing events. 
\item  A `toy' long-duration event survey, with $E=2.5 \times 10^{9}$ star years and $\epsilon(\hat{t}) = 0.4$ for $400 \, {\rm day} < \hat{t} < 
15 \, {\rm years}$ and zero otherwise, which observes no microlensing events.
\end{itemize}
For the later survey we have chosen the exposure and maximum event duration to, roughly, mimic catalogues 2 and 3 in Ref.~\cite{2018A&A...618L...4M}. The minimum event duration matches the cut-off imposed in Ref.~\cite{2022arXiv220213819B} to remove backgrounds from lensing by stars in the LMC or MW disk, and the efficiency roughly matches that obtained in their analysis.

Fig.~\ref{fig:pnobs} shows the probability distribution of the observed number of events, $P[N_{\rm obs}(f=1)]$, if all of the MW halo is in clusters containing a fixed number of PBHs for i) $M_{\rm PBH
}= 1 M_{\odot}$ and the EROS-2 like survey and ii) $M_{\rm PBH} = 10^{3} M_{\odot}$ and the `toy' long-duration event survey. For the former we consider both the EROS-2 detection efficiency, and also perfect detection efficiency, $\epsilon(\hat{t})=1$ for all $\hat{t}$.  We see that if the number of clusters in the cone to the LMC is less than of order a thousand,
the probability distribution deviates from the Poisson distribution expected for smoothly distributed DM; the peak of the distribution is shifted to a smaller value of the number of events, and there is an extended tail to large numbers of events. This behaviour is a direct consequence of the variations in the differential event rate for the shortest events discussed above. When 
the number of clusters is not large most realisations have a deficit of short events and hence a lower observed number of events than for smoothly distributed DM, while a small fraction of realisations have an excess of short events and hence a high observed number of events.  For $M_{\rm PBH} = 1 M_{\odot}$ the deviation of the probability distribution from Poissonian only emerges for $N_{\rm cl} \gtrsim 10^{6}$, much larger than the typical size of clusters for the standard PBH formation mechanism ($N_{\rm cl} \sim 10^{3}$). The deviations from the Poisson distribution are smaller for the EROS-2 detection efficiency than for perfect efficiency, because the EROS-2 efficiency is largest for $\hat{t} \approx 200$ days (and for $M_{\rm PBH} = 1 M_{\odot}$ the variations in the event duration distribution manifest at smaller values of $\hat{t}$ where the efficiency is smaller).  For $M_{\rm PBH} = 10^{3} M_{\odot}$ the deviations are visible, but relatively small, for $N_{\rm cl} =10^{3}$.

Finally we study the effect of clustering on the constraints on the fraction of the MW halo in PBHs, $f$.  For a survey which observes zero events, a $95\%$ confidence limit on the PBH halo fraction can be calculated, as in Ref.~\cite{EROS-2:2006ryy}, by finding (for each value of $M_{\rm PBH}$) the value of $f$ for which $P[N_{\rm obs}(f) =0] =0.05$. For smoothly distributed DM, $P[N_{\rm obs}(f)]$ is Poissonian and hence 
\begin{equation}
\label{poiss}
P[N_{\rm obs}(f)=0] = \exp{[-N_{\rm exp}(f)]} \,.
\end{equation}
The differential event rate is directly proportional to the local dark matter density, $\rho_{0}$, and therefore the expected number of events for smoothly distributed DM is directly proportional to $f$: $N_{\rm exp}(f) = f N_{\rm exp}(f =1)$, where $N_{\rm exp}(f=1)$ is the expected number of events for $f=1$, calculated using Eq.~(\ref{nexp}). Setting Eq.~(\ref{poiss}) equal to 0.05 gives $N_{\rm exp}(f) = 3.0$ and therefore $f = 3.0/ N_{\rm exp}(f=1)$. The constraints on $f$ for smoothly distributed DM obtained for the `EROS-2-like' survey match those found by the EROS collaboration~\cite{EROS-2:2006ryy} to within $\sim 10\%$ (e.g.~Ref.~\cite{Green:2016xgy}). For clustered DM (if 
the number of clusters within the cone is small) the probability distribution of the observed number of events is non-Poissonian, and the $95\%$ exclusion limit on $f$ has to be found by explicitly calculating $P[N_{\rm obs}(f)]$ for a range of $f$ values, to find the value of $f$ for which $P[N_{\rm obs} (f) =0] =0.05$.

For the EROS-2 like survey the change in the constraint on the halo fraction is negligible for all values of $M_{\rm PBH}$ for which $f<1$, unless $N_{\rm cl}$ is many orders of magnitude larger than expected for the standard PBH formation mechanism. For the `toy' long-duration survey the change in the constraint is only non-negligible (for physically relevant values of $N_{\rm cl}$) for large values of $M_{\rm PBH}$.  The 95\% confidence limit on the halo fraction in PBHs with $M_{\rm PBH} = 10^{3} M_{\odot}$ is $f< 0.076$ for smoothly distributed DM. For $N_{\rm cl}= 10^{3} $ the increased probability of small values of $N_{\rm obs}$ leads to a weakening of the constraint to $f< 0.096$.

\section{Discussion}
\label{sec:discuss}

We have revisited the constraints on PBH DM from stellar microlensing towards the LMC, taking into account the clustering of PBHs expected when PBHs form from the collapse of large gaussian perturbations generated by inflation. In this case the PBH clusters are sufficiently diffuse that the PBHs act as individual lenses, and clusters containing $N_{\rm cl} \sim 10^{3}$ are expected to be most abundant, with smaller clusters having evaporated. For simplicity we assume that all PBHs have the same mass, $M_{\rm PBH}$, and are in clusters containing a fixed number of PBHs, $N_{\rm cl}$. In fact some fraction of the PBHs, including those that were previously in clusters with $N_{\rm cl} \lesssim 10^{3}$, will be unclustered today, and therefore our results provide an upper limit on the effect of clustering on the LMC stellar microlensing constraints. 

We find that if the number of clusters in the cone to the LMC is sufficiently small, $ \lesssim  10^{3}$, or equivalently the number of PBHs in each cluster, $N_{\rm cl}$, is greater than $10^{6} (M_{\odot}/M_{\rm PBH})$, then the differential event rate deviates significantly from that produced by a smooth halo for short event durations. This is because the probability of there being a cluster close to the observer is small, however if there is such a cluster it produces short-duration events at a high rate. Consequently most realisations, which don't have a cluster close to the observer, have a deficit of short events (see top two rows of Fig.~\ref{fig:eventrate}). However the rare realisations where there is a cluster close to the observer have a high rate of short events (see bottom row of Fig.~\ref{fig:eventrate}). 

Consequently, as shown in Fig.~\ref{fig:pnobs}, the probability distribution of the observed number of events deviates from the Poisson distribution produced by a smooth DM distribution. It peaks at a smaller value, since most realisations have a deficit of short events, and has a long tail to large values, from the rare realisations with a cluster close to the observer which produces a high rate of short events. 
However the number of clusters is only small enough for these effects to occur if either the number of PBHs per cluster, $N_{\rm cl}$, is larger than expected, and/or the PBH mass is large.
Even for the most massive PBHs probed by decade long microlensing surveys ($M_{\rm PBH} \sim 10^{3} M_{\odot}$), the change in the constraint on the halo fraction in PBHs, $f$, is only of order ten-percent if all of the PBHs are in clusters with $N_{\rm cl} \sim 10^{3}$ (in fact not all of the PBHs are expected to be in clusters).

In summary, PBH clustering could have a significant effect on stellar microlensing constaints if the clusters are sufficiently compact (so that the cluster as a whole acts as a lens) or have a sufficiently large mass (so that the number of clusters in the cone to the LMC is small, $ \lesssim 10^{3}$). However for the most commonly studied PBH formation mechanism, the collapse of large gaussian perturbations generated by inflation, the clusters are expected to be diffuse enough that the PBHs act as lenses individually, and the number of PBHs in a typical cluster sufficiently small ($N_{\rm cl} \sim 10^{3}$), that the change in the constraints on the PBH abundance is small, even for the most massive PBHs probed by decade-long microlensing surveys.

\hspace{0.5cm}

While we were completing this work similar work by Peta\v c et al.~\cite{Petac:2022rio} appeared on the arXiv. They use a different method and take into account some effects that we neglect (e.g.~the variation of the surface density of stars in the LMC and the density profile of the PBH clusters). Nonetheless our results for the probability distribution of the observed number of events (see Fig.~\ref{fig:pnobs}) are in very good agreement with theirs. In addition we have shown that the variations in this probability distribution arise from the effect of rare clusters close to the observer on the rate of the shortest duration events. Also, following the appearance of Ref.~\cite{2022arXiv220213819B}, we have looked explicitly at the more massive PBHs, $M_{\rm PBH}$ up to $10^{3} M_{\odot}$, probed by long-duration microlensing surveys.\\


\section*{Acknowledgements}

We are grateful to Bernard Carr, Derek Inman and Mihael Peta\v c for useful discussions and/or comments. MG is supported by a United Kingdom Science and Technology Facilities Council (STFC) studentship. AMG is supported by STFC grant ST/P000703/1. For the purpose of open access, the authors have applied a CC BY public copyright licence to any Author Accepted Manuscript version arising. \\

{\bf Data Availability Statement} This work is entirely theoretical and has no associated data.

\bibliographystyle{JHEP}
\bibliography{PBH.bib}

\appendix
\section{Lensing by entire cluster}
\label{sec:app}

Appendix A2 of Ref.~\cite{Carr:2019kxo} argued that even when clusters are sufficiently extended that the PBHs act as individual lenses, lensing by the cluster as a whole renders the magnification from a single PBH unobservable. 
Here we reprise their argument, correcting the magnitude of the Einstein radius of a single PBH.

The deflection angle for light which `grazes' the radius of a cluster is
\begin{equation}
\label{alpha}
\alpha = \frac{4 GM_{\rm cl}}{c^2 R_{\rm cl}} 
       \approx  5 \times 10^{-11} \left( \frac{M_{\rm cl}}{10^{3} M_{\odot}} \right) \left( \frac{ 4 \, {\rm pc}}{R_{\rm cl}} \right) \,,
\end{equation}
where we have normalised (roughly) to the mass and radius of the smallest clusters that will not have evaporated by the present day ($N_{\rm cl} \sim 10^{3}$) if $M_{\rm PBH}= 1 M_{\odot}$. This small deflection isn't observable, however the light from the lensed star will be spread over an arc with length $l \sim x L \alpha$. Ref.~\cite{Carr:2019kxo} argues that if $M_{\rm cl} \gtrsim 10^{3} M_{\odot}$ then this arc length is much larger than the Einstein radius of an individual Solar mass PBH, $l \gg R_{\rm E}$. Therefore the star's luminosity will only be marginally affected by lensing by an individual PBH and hence a `classic' microlensing event will not occur. However they assume $R_{\rm E} \sim 10^{-8} \, {\rm pc}$. Using Eq.~(\ref{alpha}) and Eq.~(\ref{re}), which gives $R_{\rm E} \sim 10^{-4} \, {\rm pc}$ for $M_{\rm PBH} = 1 M_{\odot}$, we find
\begin{eqnarray}
\frac{l}{R_{\rm E}} &\approx& 0.02 \left( \frac{x}{1-x} \right)^{1/2} \left( \frac{M_{\rm cl}}{10^{3} M_{\odot}} \right) \nonumber \\
&& \times  \left( \frac{4 \, {\rm pc}}{R_{\rm cl}} \right) 
\left( \frac{L}{50 \, {\rm kpc}} \right)^{1/2} \left( \frac{M_{\odot}}{M_{\rm PBH}}\right)^{1/2}   \,, 
\end{eqnarray}
and using Eq.~(\ref{rcl}) and $M_{\rm cl} = N_{\rm cl} M_{\rm PBH}$,
\begin{eqnarray}
\frac{l}{R_{\rm E}} &\approx& 0.02 \left( \frac{x}{1-x} \right)^{1/2} \left( \frac{N_{\rm cl}}{10^{3}} \right)^{1/6}  \nonumber \\
&& \times \left( \frac{L}{50 \, {\rm kpc}} \right)^{1/2} \left( \frac{M_{\rm PBH}}{M_{\odot}} \right)^{1/6} \,.
\end{eqnarray}

\end{document}